# GLOBAL DYNAMICAL MODEL OF THE CARDIOVASCULAR SYSTEM

**Alexander S. Kholodov**[1], **Sergey S. Simakov**[2], **Yaroslav A. Kholodov**[3],
**Alexey A. Nadolskiy**[4], **Alexander N. Shushlebin**[5]

[1-3] Moscow Institute of Physics and Technology
9, Instituskii Lane, Dolgoprudny, Moscow Region, Russia, 141700
e-mail: [1] xolod@crec.mipt.ru, [2] simakov@crec.mipt.ru, [3] kholodov@crec.mipt.ru

[4-5] All-Russian Institute of Technical Physics
P.O.Box 245, Snezhinsk, Chelyabinsk Region, Russia, 456770
[4] a.a.nadolsky@vniitf.ru, [5] a.n.shushlebin@vniitf.ru

**Keywords:** Cardiovascular Mechanics, Global Circulation, Loss of Blood.

*Abstract. Blood system functions are very diverse and important for most processes in human organism. One of its primary functions is matter transport among different parts of the organism including tissue supplying with oxygen, carbon dioxide excretion, drug propagation etc. Forecasting of these processes under normal conditions and in the presence of different pathologies like atherosclerosis, loss of blood, anatomical abnormalities, pathological changing in chemical transformations and others is significant issue for many physiologists. In this connection should be pointed out that global processes are of special interest as they include feedbacks and interdependences among different regions of the organism. Thus the main goal of this work is to develop the model allowing to describe effectively blood flow in the whole organism. As we interested in global processes the models of the four vascular trees (arterial and venous parts of systemic and pulmonary circulation) must be closed with heart and peripheral circulation models. As one of the model applications the processes of the blood loss is considered in the end of the paper.*

A.S. Kholodov, S.S. Simakov, Y.A. Kholodov, A.A. Nadolskiy, A.S. Shushlebin

# 1 INTRODUCTION

Investigations of the blood flow mechanics were started long time ago. Most likely Euler in 1775 was the first who wrote equations for the blood flow in arteries. In addition, Young was the first who calculated the pulse wave velocity [1] in the artery of a human. Womersley summarized the physics of pulsative flow in the channels having flexible walls [2, 3] and others. However, great numbers of nowadays publications affirm urgency, importance and incompleteness of the problem of cardiovascular system modeling.

Often in biomechanics, the blood is considered as a fluid. In simplest cases it supposed to be single-component, non-viscous, non-compressible while the most complicated models include chemical reactions between the components dissolved in blood. In any case, it should be mentioned that blood has very complicated rheological properties. It may be considered in terms of continuum media due to the certain conditions taken place in most parts of the circulatory system of the organism under the normal conditions [4].

Wide range of the proposed models may be classified by their dimensionality. Two-dimensional [5, 6] and three-dimensional [7, 8] models are widely used for the blood flow analysis in specific part of the circulatory system. If applied to the tasks of global circulation such approaches require huge computational resources. One-dimensional models [4, 9-16] are often exploited for investigations of the pulse wave propagation and reflection from the vessel bifurcation. Such approach is more effective for the tasks of global circulation but it requires identification of the great number of ill-conditioned and strongly variable parameters describing the model. It also exist lumped models [7, 17-19] that in one of the most known cases are based on electro-mechanical analogies [20]. The latest approaches are based on the multiscale technique that combines the models of different dimensionality in order to include possibly greater parts of circulatory system e.g. [7].

In fact, circulatory system is a single whole structure that results in non-linear internal effects. Thus, ideal model must include possibly greater parts of circulatory system. In addition, such model allows to simulate different global processes such as feedbacks, interdependences among different regions of the circulatory system, matter transfer and others.

As a basis for such model, we propose to use the model of non-linear pulsative flow of viscous incompressible fluid streaming through the collapsible tube [14-16]. The flow is supposed to be pseudo one-dimensional as all values are supposed to be averaged over cross-sectional area of the vessel. By means of specific boundary conditions it may be generalized for the case of the pulsative flow through the graph of the collapsible tubes. Every edge of such graph corresponds to the vessel in cardiovascular system. Every node corresponds to the vessels bifurcation. The heart in these terms corresponds to the one specific node and performs pumping function. The other group of specific nodes must be considered in the graph that corresponds to the junction points with microcirculation. In fact peripheral circulation is hard to describe in terms of such model. It requires to determine dramatically huge number of the parameters and computational resources. Thus some averaging procedures required to produce virtual vessels having properties corresponding to some macro area of the capillary channel. Another way to deal with this problem is to include the model of peripheral circulation based on the principle of liquid filtration through the porous media. The multiscale approach can be realized using this model by substituting some part of the vascular network with appropriate local 3D model.

To the knowledge of the authors quite a few reconstructions of the whole circulatory system using at least 1D approach are exist at present [13, 21]. In addition linear theory was used in some approaches that simplified analytical and computational analysis and allows obtaining experimentally proven results but it still carried some inaccuracy in the model.

## 2   MATHEMATICAL BASIS

Physically blood flow is as pulsatile flow of incompressible fluid streaming through the network of the vessels. As the base model for such flow, we propose to use pseudo one-dimensional model of non-stationary incompressible fluid streaming through the collapsible tube. All such tubes must be connected each other by the appropriate boundary conditions.

For every vessel of $k^{th}$ generation the mass and momentum conservation are [4, 9-16]:

$$\partial S_k / \partial t + \partial (u_k S_k) / \partial x = \varphi_k(t, x, S_k, u_k, r_i) \quad (1)$$

$$\partial u_k / \partial t + \partial (u_k^2/2 + p_k/\rho_k) / \partial x = \psi_k(t, x, S_k, u_k, r_i) \quad (2)$$

where $t$ — time; $x$ — distance counted from the junction point with vessel of younger generation; $\rho$ — blood density; $k$ — index of the vessel; $S_k(t,x)$ — cross-sectional area; $u_k(t,x)$ — linear velocity of the flow averaged over cross-section; $p_k(t,x)$ — pressure in the vessel counted off from atmospheric; $\psi_k$ — the impact from the external forces (gravitation, friction and others); $\chi_{ki}$ — parameters describing the impact $i$ on the vessel $k$.

As the flow described by the (1),(2) is supposed to be in collapsible tube having thin walls elastic properties of the walls must be specified. This relation may be regarded as "equation of state" for the wall of the vessel:

$$p_k - p_{*k} = \rho c_k^2 f_k(S_k) \quad (3)$$

where $c_k$ — rate of small disturbance propagation; $p_{*k}$ — pressure in the tissues surrounding the vessel. For the most vessels the value of $p_*$ is supposed to have zero value and just for the small pulmonary vessels it supposed to be equal to the pleural pressure induced by the bronchial tubes of respiratory system. The specific form of the state equation (3) depends on the type and size of the vessel [4, 10]. In this work for the most vessels was used the following:

$$f_k(S_k) = \begin{cases} \exp(S_k/S_k^0 - 1) - 1, & S_k > S_k^0 \\ \ln(S_k/S_k^0), & S_k \le S_k^0 \end{cases} \quad (4)$$

and for some cases:

$$f_k(S_k) = \begin{cases} (S_k - 1)/\alpha, & S_k > S_k^0 \\ \beta/S_k^{3/2}, & S_k \le S_k^0 \end{cases} \quad (5)$$

where $S_{0k}$ — mean cross-sectional area averaged over one cardiac cycle; $\alpha = 0.204\, kPa^{-1}$ for arteries and $\alpha = 0.068\, kPa^{-1}$ for veins; $\beta = 2.5\, kPa$.

In order to state correct boundary conditions for the task (1)-(3) the characteristic form is required that can be rewritten using the following vectorial notations:

$$\vec{V}_k = \{S_k, u_k\}, \quad \vec{F}_k = \{u_k S_k, p_k/\rho_k + u_k^2/2\} \text{ and } \vec{g}_k = \{\varphi_k, \psi_k\}$$

along the $i^{th}$ characteristic:

$$\vec{W}_{ki} \cdot d\vec{V}_k / dt = \vec{W}_{ki} \cdot (\partial \vec{V}_k / \partial t + \lambda_{ki} \partial \vec{V}_k / \partial x) = \vec{W}_{ki} \cdot \vec{g}_k, \quad i = 1, 2 \quad (6)$$

where $\lambda_{ki}$ — eigenvalues of Jacobi's matrix $A_k = \partial \vec{F}_k / \partial \vec{V}_k$ obtained from the equation:

$$Det(A_k - \lambda_k E) = 0 \tag{7}$$

where $E$ — unit matrix; $\vec{W}_{ki}$ — left eigenvectors of matrix $A_k$ that obtained from the set:

$$\vec{W}_{ki} \cdot (A_k - \lambda_{ki} E) = 0 . \tag{8}$$

From the equations (1–3) it follows ($i = 1, 2$):

$$\lambda_{ki} = u_k + (-1)^{i+1} c_k \sqrt{S_k \frac{\partial f_k}{\partial S_k}}; \ \vec{W}_{ki} = \left\{ c_k \sqrt{\frac{\partial f_k}{\partial S_k}}, \ (-1)^{i+1} \sqrt{S_k} \right\} . \tag{9}$$

Equation (6) must be combined with the other boundary conditions stated at the entry and terminal points of every vessel. In this connection if entry point considered the second characteristic curve $(i = 2)$ is chosen in (9) that then substituted to (6). At the same time if terminal point is under the discussion the first characteristic curve $(i = 1)$ is chosen. Keeping this observation in mind the rest part of the boundary conditions can be stated.

At the internal points of every vessel (that are not entry and terminal points), the initial conditions can be chosen rather arbitrary, e.g.:

$$S_k(0, x) = S_k^0, \ u_k(0, x) = Q_0 / S_k^0, \ k = 1, ..., K \tag{10}$$

After that, several cardiac circles must be simulated before stationary conditions corresponding to the norm occur.

At the entry points of the vessels (*I*) connected to the heart ($x = 0$) the blood flow is assigned as the boundary condition:

$$u_I(t, 0) \cdot S_I(t, 0) = Q_I(t, 0) \tag{11}$$

At the terminal points from the venous system $x = X_{K_v}$ the pressure is set as the boundary condition:

$$p_{K_v}(t, X_{k_v}) = p_v(t) . \tag{12}$$

As global cardiovascular system considered the values of $Q_I(t, 0)$ and $p_v(t)$ must be obtained from the set of equation describing the heart functioning [14-16].

At the bifurcations of the vessels the condition of the pressure difference (Poiseuille's law) for every branch $k = k_1, ..., k_M$ composing the node:

$$p_k(t, x) - p_l(t) = \varepsilon_k R_k Q_k(t, x), \ k = k_1, k_2, ..., k_M \tag{13}$$

and the mass conservation condition:

$$\sum_{k = k_1, ..., k_M} \varepsilon_k Q_k(t, x) = \sum_{k = k_1, ..., k_M} \varepsilon_k u_k S_k = 0 \tag{14}$$

are stated. Here $R_k$ designates the resistance of $k^{th}$ branching section. For the branches incoming into a node: $\varepsilon_k = 1$, $x = X_k$ (terminal point) while for the branches outgoing from a node: $\varepsilon_k = -1$, $x_k = 0$ (entry point).

The blood flow in the region of the vessel bifurcation has very complicated structure. Therefore, such flow cannot be considered in terms of one-dimensional model and at least two-dimensional approach should be used instead. Such approach would resulted in dramati-cally increase of the computational cost because the thousands of the bifurcations have to be considered in our model. In order to simplify the model the equations (13), (14) together with (6) was proposed. Such approach seems to be reasonable in this case as the specific dimension of the bifurcation region is rather small (it has the same order as the largest diameter of the vessels in bifurcation) in comparison with the length of the vessel. Actually the more accurate approach must include the limit ratios from (1), (2) at every vessel bifurcation. They can be calculated from the Bernoulli's equation in this case [13]. Such simulations were carried out. The results show that in the most cases the influence of the equation members proportional to the squared velocity is negligible in comparison with the members proportional to the pressure in the region of bifurcation. This fact is also consistent with the analytical analysis using linear approach [21, 22].

At the junction points with the peripheral circulation the condition similar to (13) is used. It connected with the task of combining the models of peripheral circulation [23] and the model of the blood flow in large vessels [15, 16].

For numerical solving of the equations similar to (1), (2) there are many various numerical methods. However, special investigation was required and set of numerical methods developed before [24-27] was used to design software implementation of the model. In this connection one of the main difficulty was that the dimensions of the small vessels are differ from that for the large vessels by the factor of $10^2 \div 10^3$ that impose quite a strict constrains to the time step. The proposed methods allow to overcome this problem.

## 3   RESULTS

The one-dimensional structure of the vessels was reconstructed basing on the known anatomical data [28, 29]. Specially developed software was used to convert them to the format suitable for computational simulations. The parameters of the models were identified from the anatomical and physiological data [4, 28-30] and some parameters were chosen basing on the model output to fit it with known data. Some simulations related to the parameters identification were also presented at [14-16].

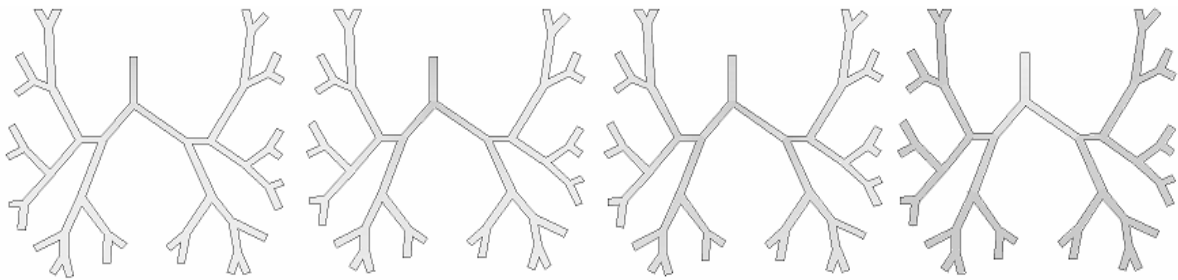

Figure 1: Pressure wave propagation through the arteries of pulmonary circulation.

The results of the blood flow simulations in the arterial part of pulmonary and systemic circulation under the normal conditions are represented in fig.1 and fig.2 correspondingly. Grayscale designates divergence from the minimum value of the pressure. The scale is chosen separately for every vessel to make obvious the pulse wave propagation. Flow patterns during different distinctive parts of the period of one cardiac cycle are depicted from left to right.

As a result, simulations reveal quite satisfactory correlation between the calculated data and the results of the experiments obtained from the known literature. It allows to state adequateness of used mathematical approaches to the model construction as well as correct identification of the structural and functional properties of the cardiovascular system.

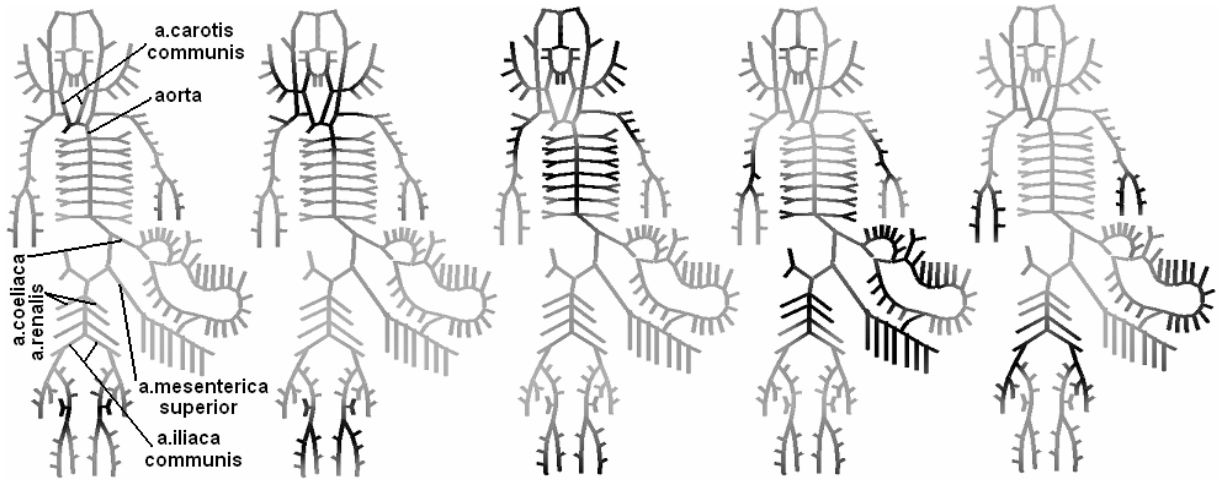

Figure 2: Pressure wave propagation through the arteries of systemic circulation.

As one of the models application, the loss of blood during the damage of the common carotid artery is considered in this work. In terms of the proposed model this process can be simulated by the appropriate right-hand member in (1):

$$\varphi_K = -\alpha \cdot S_K(t, x_*) \qquad (15)$$

where $K$ — index of the common carotid artery; $x_*$ — coordinate of the damage from the entry to the artery; $\alpha$ — coefficient of the blood loss intensity.

Computational experiments depicted in the fig. 3, 4 show the pressure and velocity patterns as under the normal conditions (*a*) and during the damage of the common carotid artery (*b-d*). The following coefficients of the blood loss intensity were taken for these simulations: $\alpha_b = 5 \; ml/\sec$; $\alpha_c = 10 \; ml/\sec$; $\alpha_d = 15 \; ml/\sec$.

From fig. 3 it is easy to observe that due to the loss of blood the flow velocity is heavily decreased right ahead the damaged area and increased up to the normal value at the exit of the vessel. At the *b-d* parts of the fig. 4 the pressure decrease before the damaged area can be observed along with its increase after the area of outflow. Such changes in the dynamic of the blood flow demonstrate that the flow pattern during the damage of the vessel tends to return to its normal conditions at the entry and terminal points. The changes in the blood flow during the damage are also result in blood afflux to the area of injury and therefore in increasing concentration of the substances that give rise to the blood coagulation. The simulations reveal that both effects may be explained just by the elastic properties of the vessel walls without any additional regulation.

It must be pointed out that mentioned numerical experiments dealt with local process as they consider just one vessel during one heart cycle. The further experiments will examine changes in blood flow taking place in the whole cardiovascular system during the loss of blood having different duration and intensity at the local areas.

A.S. Kholodov, S.S. Simakov, Y.A. Kholodov, A.A. Nadolskiy, A.S. Shushlebin

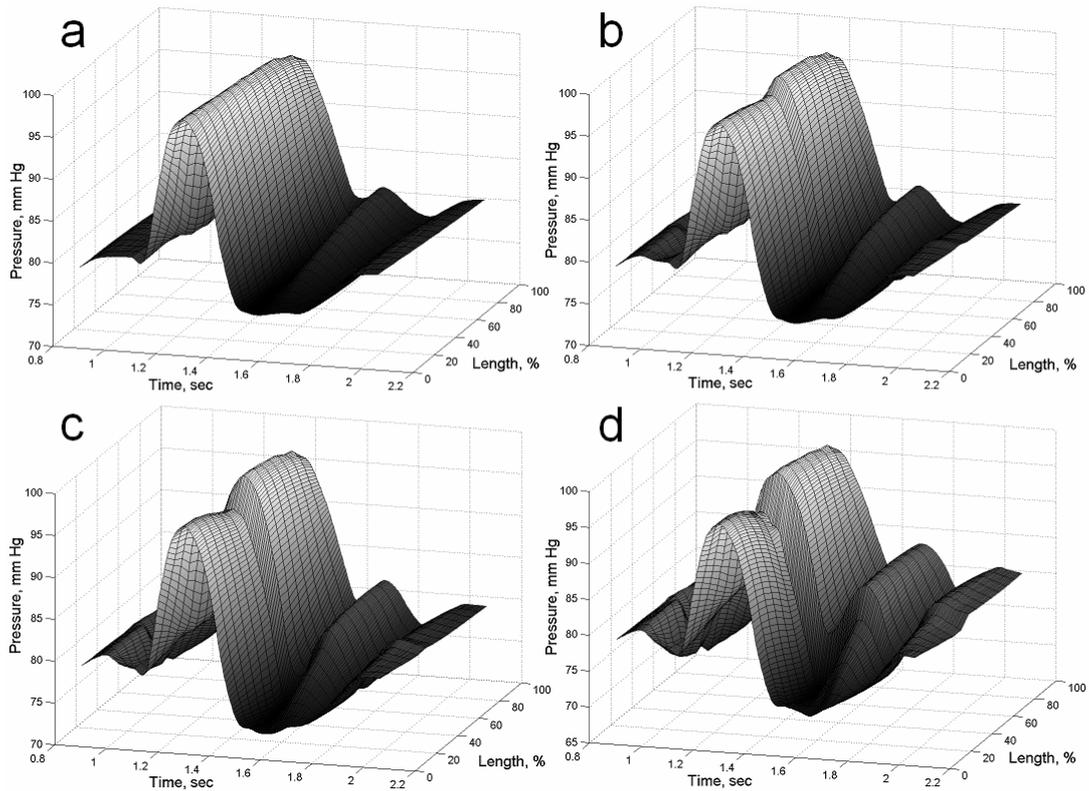

Figure 3: Pressure of the blood in common carotid artery under the normal conditions (a) and during the loss of blood in the center of the vessel (b-d).

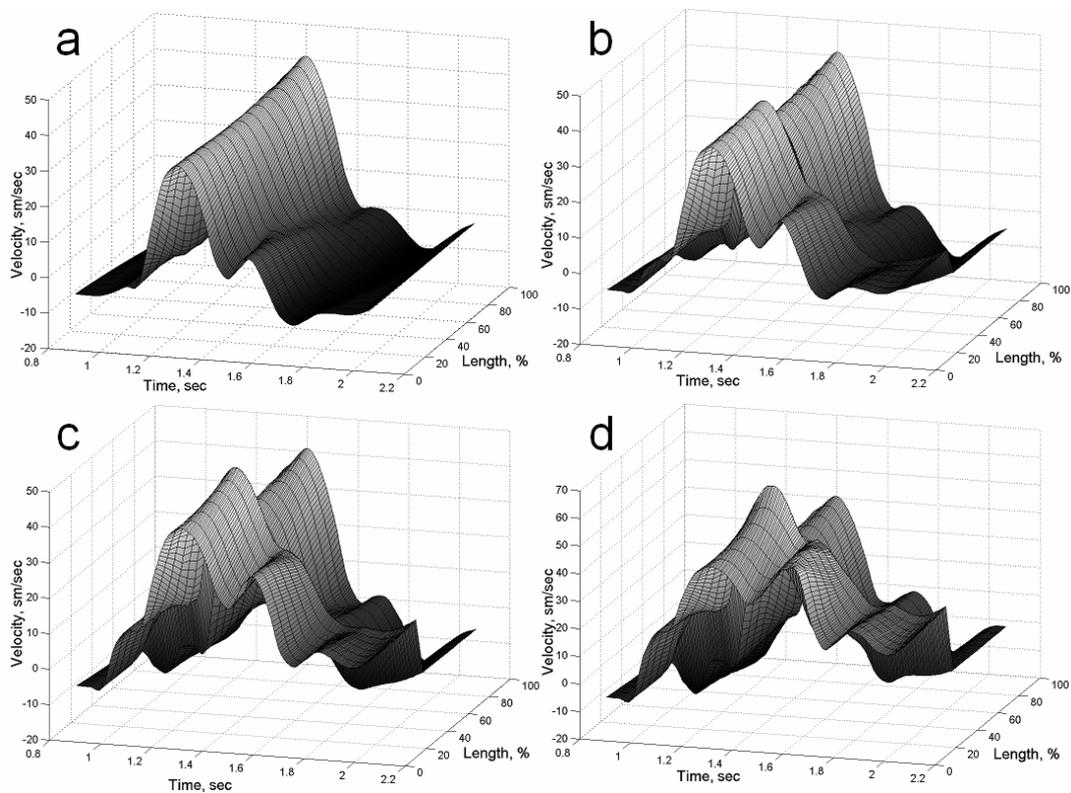

Figure 4: Linear velocity of the blood flow in common carotid artery under the normal conditions (a) and during the loss of blood in the center of the vessel (b-d).

## 4   CONCLUSIONS

- Basing on the analysis of the existing methods of cardiovascular system modeling an approach was proposed that allow to simulate cardiovascular system as a whole. Constituent parts of the proposed model may be detailed, averaged or substituted by 2D or 3D elements thus allowing multiscaling.

- The carried out simulations show flow pattern in the arteries of the pulmonary and systemic circulations. They also reveal that due to the elastic properties of the vessels during the short-time loss of blood the flow pattern tends to return to its normal state.

- The proposed model of global circulation is effective and useful for practical applications. It may be extended and adapted for many different tasks of computational biomechanics.

**REFERENCES**

[1] Young T., On the function of the heart and arteries: The croonian lecture. *Phil. Trans. Roy. Soc*, 99, 1-31, 1809.

[2] Womersley J.R., Velocity profiles of oscillating arterial flow with some calculations of viscous drag and the Reynolds number. *J. Physiology*, 128, 629-640, 1955.

[3] Womersley J.R., Oscillatory flow in arteries: the constrained elastic tube as a model of arterial flow and pulse transmission. *Physics in Medicine and Biology*, 2, 178-187, 1957.

[4] Caro C.G., Pedley T.J., Schroter R.C., Seed W.A., *The mechanics of the circulation*, Oxford University Press, New York, Toronto, 1978.

[5] Chakravarty S., Mandal P.H., A nonlinear two-dimensional model of blood flow in an overlapping arterial stenosis subjected to body acceleration. *Math. Comput. Modeling*, **24**(1), 43-58, 1996.

[6] Yakhot A., Grinberg L., Nikitin N., Modeling rough stenosis by an immersed-boundary method. *J. Biomechanics*, **38**(5), 1115-1127, 2005.

[7] Logana K., Balossino R., Migliavacca F. et. al., Multiscale modeling of the cardiovascular system: application to the study of pulmonary and coronary perfusion in the univentricular circulation. *J. Biomechanics*, **38**(5), 1129-1141, 2005.

[8] Peskin, C.S., McQueen, D.M., et.al., Fluid dynamics of the heart and its valves. *Case studies in mathematical modeling, ecology, physiology, and cell biology*, Prentice-Hall, Inc., Englewood Cliffs, New Jersey, chapter 14, 309-337, 1996.

[9] Voltairas P.A., Fotiadis D.I., et. al., Anharmonic analysis of arterial blood pressure and flow pulses. *J. Biomechanics*, **38**(7), 1423-1431, 2005.

[10] Olufsen, M., Structured tree outflow condition for blood flow in larger systemic arteries. *Am. J. Physiol.*, **276**(1), H257-H268, 1999.

[11] Pedley, T.J., Mathematical modelling of arterial fluid dynamics. *J. Engng. Math.*, 47, 419-444, 2003.

[12] Van de Vosse, F.N., Mathematical modelling of the cardiovascular system. *J. Engng. Math.*, 47, pp. 175-183, 2003.

[13] Abakumov M.V., Gavrilyuk K.V., Favorskii A.P., et al., Mathematical Model of Hemodynamics of Cardiovascular System. *J. Differential. Equations*, **3**(7), 892–898, 1997.

[14] Kholodov A.S., Some dynamical models of external breathing and blood circulation taking into account their interaction and matter transfer. *Mathematical models and medicine progress*, Nauka, Moscow, 121-163, 2001.

[15] Kholodov A.S., Simakov S.S., Evdokimov A.V., Kholodov Y.A., Numerical simulations of cardiovascular diseases and global matter transport. *Proc. of the Int. Conf. Advanced Information and Telemedicine Technologies for Health*, Ablameyko S. et. al. eds., Minsk, **2**, 188–192, 2005.

[16] Kholodov A.S., Simakov S.S., Evdokimov A.V., Kholodov Y.A., Matter transport simulations using 2D model of peripheral circulation coupled with the model of large vessels. *Proc. of II Int. Conf. On Comput. Bioeng.*, Rodrigues H. et. al. eds., IST Press, 1, 479-490, 2005.

[17] Burnette R.R., Computer simulation of human blood flow and vascular resistance. *Comput. Bio. Med.*, **26**(5), 363-369, 1996.

[18] Byczkowski J.Z., A linked pharmacokinetic model and cancer risk assessment for breast-fed infants. *J. Drug Info.*, 30, 401-412, 1996.

[19] Godfrey K., *Compartmental models and their application*, Academic Press, London, 1983.

[20] Olufsen M., Nadi A., On deriving lumped models for blood flow and pressure in the systemic arteries. *Math. biosciences and engineering*, **1**(1), 61-80, 2004.

[21] Ashmetkov I.V., Favorskii A.P., et.al., Methodology of Mathematical Modeling of cardiovascular system. *Math. Mod.*, **12**(2),106-117, 2000.

[22] Ashmetkov I.V., Favorskii A.P., et. al. Boundary problem for linearized equations of hemodynamics on graph. *Differential Equations*, **40**(1), 1–11, 2004.

[23] Evdokimov A. V., Kholodov A. S., Pseudo-steady Spatially Distributed Model of Human Blood Circulation. *Computer Models and Medicine Progress*, Nauka, Moscow, 164–193, 2001

[24] K.M. Magomedov, A.S. Kholodov, *Grid-characteristic numerical methods*. Nauka, Moscow, 1988.

[25] O.V. Vorobyov, Y.A. Kholodov, On the method of numerical integration of 1D problems in gas-dynamics. *J. Math. Mod.*, **8**(1), 77-92, 1996.

[26] A.S. Kholodov, Monotonic difference schemes on irregular grids for elliptic equations in domains with multiple boundaries. *J. Math. Mod.*, **3**(9), 104-113, 1991.

[27] A.S. Kholodov, About majorized differential schemes on irregular grids for hyperbolic equations. *Math. Mod.*, Moscow, MSU, 105-113, 1993.

[28] *Gray's Anatomy: The Anatomical Basis of Medicine and Surgery*, Churchill-Livingstone, 2004,

[29] Vorobiyov V.P., *The big atlas of human's anatomy*. Harvest, 2003.

[30] R.F. Schmidt, G. Thews eds., *Human Physiology*. Mir, Moscow, **2**, 1996.